\documentclass[final,prl,twocolumn,amsmath,amssymb,amsfonts,showpacs,titlepage,nobalancelastpage,raggedbottom,floatfix]{revtex4}

\usepackage[english]{babel}
\usepackage{ifpdf}
\ifpdf
\usepackage[pdftex]{graphicx}
\usepackage[protrusion=true,expansion=true]{microtype}  
\usepackage[pdftex,colorlinks,linkcolor=black,citecolor=black,urlcolor=black,hyperfootnotes=false]{hyperref}
\else
\usepackage[dvips]{graphicx}
\usepackage{microtype}
\fi

\newcommand{\integral}[4]{
\int_{#3}^{#4}\mathrm{d}#2 \, #1 }

\begin{document}
\selectlanguage{english}
\title{Survival of the Aligned: Ordering of the Plant Cortical Microtubule Array}
\author{Simon H. Tindemans}
\author{Rhoda J. Hawkins}
\altaffiliation[Current address: ]{UMR 7600, UPMC
/CNRS, 4 Place Jussieu, 75255 Paris Cedex 05 France}
\author{Bela M. Mulder}
\affiliation{FOM Institute AMOLF, Science Park 104, 1098 XG,
Amsterdam, The Netherlands}
\begin{abstract}
The cortical array is a structure consisting of highly aligned microtubules which plays a crucial role in the
characteristic uniaxial expansion of all growing plant cells. Recent experiments have shown polymerization-driven
collisions between the membrane-bound cortical microtubules, suggesting
a possible mechanism for their alignment. We present both a coarse-grained theoretical model and stochastic particle-based simulations
of this mechanism, and compare the results from these complementary approaches. Our results indicate that collisions
that induce depolymerization are sufficient to generate the alignment of microtubules in the cortical array.
\end{abstract}

\pacs{87.16.Ka, 87.16.ad, 87.16.af, 87.16.Ln}

\maketitle


Microtubules are a ubiquitous component of the cytoskeleton of eukaryotic cells. These dynamic filamentous protein
aggregates, in association with a host of microtubule associated proteins (MAPs), are able to self-organize into
dynamic, spatially extended stable structures on the scale of the cell \cite{Alberts}.
\begin{figure}[ht!]
\begin{center}
\includegraphics{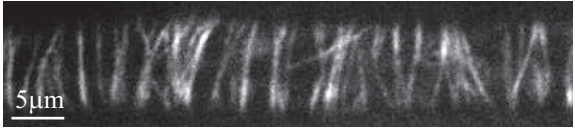}
\caption{Transverse cortical array in an etiolated dark-grown
\emph{Arabidopsis thaliana} hypocotyl cell with fluorescently
labeled microtubules. Image courtesy of Jelmer Lindeboom,
Wageningen University.}\label{figure1}
\end{center}
\end{figure}
In contrast to the more commonly studied animal cells, plant cells
are encased in a cellulosic cell wall, and generally only expand
along a single well-defined growth axis. A crucial component in this
anisotropic growth process is a plant-unique microtubule structure
called the cortical array \cite{EhrhardtShaw06}. This structure
consists of highly aligned microtubules attached to the inner side
of the cell membrane and oriented transversely to the growth
direction (see Fig.\ 1) and establishes itself in a period of about
one hour after cell division. The cortical array has two particular
features, both related to the fact that the microtubules are bound
to the cell membrane \cite{Shaw+03,Vos+04}: (i) it is effectively a
2-dimensional system and (ii) the cortical microtubules do not slide
along the membrane, so the only displacements are caused by the
ongoing polymerization and depolymerization processes intrinsic to
microtubules. As a consequence of these two constraints, the
so-called plus end of a growing cortical microtubule can `collide'
with another microtubule. Recent experiments \cite{DixCyr04} have
shown that these collisions indeed occur and can have three possible
outcomes whose relative frequency is determined by the angle between
the microtubules involved (see Fig.\ \ref{figure2}a). The first
option is that the incoming microtubule changes its direction and
continues to grow alongside the microtubule it encountered, an
outcome that is predominant at smaller angles and is known as
`zippering'. The second option is the so-called `induced
catastrophe', in which the incoming microtubule switches to the
shrinking state. Finally, there is a possibility that the incoming
microtubule simply `crosses over' the obstacle, continuing to grow
in its original direction.

In this Letter we address the question of whether, as has been posited by  Dixit and Cyr \cite{DixCyr04}, these
interactions are sufficient to explain the alignment of microtubules in the cortical array. To do so we construct a
model for the microtubule dynamics and interactions, and evaluate it using two complementary approaches: a
coarse-grained theory and particle-based simulations. The theory allows us to reduce the size of the model parameter
space by identifying the relevant control parameter of the system and establishes the criteria for spontaneous
symmetry breaking to occur. The simulations explicitly consider the stochastic dynamics of individual microtubules,
and are thereby able to test the validity of the theory. The simulations can also be extended to include known other
contributing effects such as minus-end treadmilling and microtubule severing proteins, but here we focus on a
minimal version of the model that can be addressed using both the theoretical and simulation approaches in order to
establish a reference system and test the general hypothesis of \cite{DixCyr04}.

Our model differs from existing models for 2D organization of filamentous proteins in two important ways. Firstly,
in most of these models the filaments are both free to rotate and translate as a whole
\cite{geigant98,Zum+05,kruse05,aranson06,ruehle08}, which is inconsistent with the experimental observations on the
cortical array. Secondly, our model explicitly takes into account the dynamic instability of the individual
microtubules, providing the potential for intrinsic stabilization of the microtubule length distribution. This
differs from the model by Baulin \emph{et al}.\ \cite{Baulin2007} in which deterministically elongating microtubules stop growing only while
obstructed by other microtubules. The lack of an intrinsically bounded
length most likely precludes the existence of stable stationary states in their simulations.

For the intrinsic microtubule dynamics in our model, we use the
standard two-state dynamic instability model \cite{Marileen93} in
which each microtubule plus end is assumed to be either growing with
a speed $v^{+}$ or shrinking with a speed $v^{-}$. This plus end can
switch stochastically from growing to
shrinking (a so-called `catastrophe') with rate $r_{\textrm{c}}$, or from shrinking to growing (a so-called `rescue') with rate $%
r_{\textrm{r}}$ in a process known as dynamic instability. New
microtubules are nucleated isotropically and homogeneously with a
constant rate $r_{\textrm{n}}$. The microtubule minus ends are
assumed to remain attached to their nucleation sites.

\begin{figure}[hb!]
\begin{center}
\includegraphics[width=0.4\textwidth]{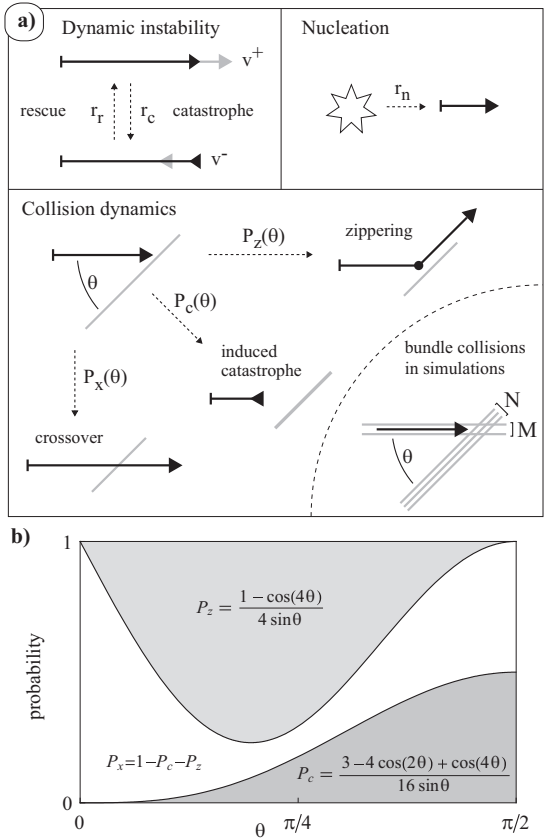}
\caption{a) schematic overview of the included effects and parameters in the model. b) relative frequency
of collision outcomes as a function of angle of incidence used in our model.}\label{figure2}
\end{center}
\end{figure}

Because the persistence length $l_p$ of microtubules is long ($\sim \mathrm{mm}$) compared to the average length of
a microtubule ($\sim 10\mu\mathrm{m}$) and thermal motion is inhibited by the attachment to the plasma membrane,
microtubules are modelled as straight rods with kinks at positions where a zippering event has occurred. A
microtubule therefore consists of a series of connected segments to which we assign an index $i$, starting at $i=1$
for the segment attached to the nucleation site. In light of the available evidence, we assume that the
angle-dependent collision outcome probabilities $P_{\textrm{z}}$ (zippering), $P_{\textrm{c}}$ (induced catastrophe)
and $P_{\textrm{x}}$ (crossover) are independent of the polarity of the microtubules and are therefore fully defined
on the interval $[0,\frac{\pi}{2}]$.

We first analyze this system using a coarse-grained theory, in
which we consider densities of microtubule segments instead of
individual microtubules. From the outset we assume the system is,
and remains, spatially homogeneous, and later restrict ourselves
to steady-state solutions. Because microtubules are nucleated
isotropically and can change their orientation after each
zippering event, we introduce separate densities for each segment
index $i$. Furthermore, length changes and collisions can only occur in
segments that contain the microtubule plus end. Therefore, we
further distinguish the \emph{active} segments, containing either
a growing (+) or shrinking (-) plus end, and the \emph{inactive}
(0) segments that form the `body' and tail of the microtubule. Our
variables are thus the areal number densities $m_{i}^{\sigma
}(l,\theta ,t)$ of segments in state $\sigma \in \left\{
0,-,+\right\} $ with segment index $i$, having length $l$ and
orientation $\theta $ (measured from an arbitrary axis) at time
$t$. From these, we compute the total length density $k(\theta,t)$
as
\begin{equation}
k(\theta,t) = \sum_i \negthickspace \int_0^{\infty }\negthickspace \mathrm{d} l \; l \left[ m_i^+(l,\theta,t) +
m_i^-(l,\theta,t) + m_i^0 (l,\theta,t)\right] .
\end{equation}
The segment densities obey a set of evolution equations that can
symbolically be written as
\begin{subequations}\label{eq:MEall}
\begin{align}
\partial _{t}m_{i}^{+}(l,\theta,t) =&\Phi_{\text{\textit{grow}}}[m_{i}^{+}]+ \Phi_{\text{\textit{rescue}}}[m_{i}^{-}]-\Phi
_{\text{\textit{sp.cat.}}}[m_{i}^{+}]\nonumber\\&-
\Phi_{\text{\textit{ind.cat.}}}[m_{i}^{+},k]-\Phi
_{\text{\textit{zip}}}[m_{i}^{+},k]
\label{eq:MEgrow} \\
\partial _{t}m_{i}^{-}(l,\theta,t) =&\Phi
_{\text{\textit{shrink }}}[m_{i}^{-}]-\Phi
_{\text{\textit{rescue}}}[m_{i}^{-}]+\Phi
_{\text{\textit{sp.cat.}}}[m_{i}^{+}]\nonumber\\&+\Phi
_{\text{\textit{ind.cat.}}}[m_{i}^{+},k] +\Phi
_{\text{\textit{react.}}}[m_{i}^{+},m_{i+1}^{-},k]
\label{eq:MEshrink} \\
\partial _{t}m_{i}^{0}(l,\theta,t) =& \Phi
_{\text{\textit{zip}} }[m_{i}^{+},k] -\Phi
_{\text{\textit{react.}}}[m_{i}^{+},m_{i+1}^{-},k]
\label{eq:MEinactive}
\end{align}
\end{subequations}
The arguments in square brackets explicitly display the functional dependencies of the terms on the right hand side.
Below, we explain each of these terms briefly, and refer the reader to \cite{Hawkins09} for a full derivation and an
in-depth analysis. The dynamics of the active growing ($+$) and shrinking ($-$) segments of microtubules unperturbed
by interactions are given by the standard spontaneous catastrophe and rescue rates
$\Phi_{\text{\textit{sp.cat.}}}[m^{+}]=r_\textrm{c} m^{+}$ and $\Phi_{\text{\textit{rescue}}}[m^{-}]=r_\textrm{r}
m^{-}$, and the advective terms $\Phi_{\text{\textit{grow}}}[m^+]= - v^+ \frac{\partial m^{+}}{\partial l}$ and $\Phi
_{\text{\textit{shrink}}}[m^{-}]= v^- \frac{\partial m^{-}}{
\partial l}$ due to growth and shrinkage respectively
\cite{Marileen93}. Collisions between microtubules that lead to an induced catastrophe cause growing segments to
switch to the shrinking state, at a rate given by $\Phi_{\text{\textit{ind.cat.}}}[m^{+},k]=v^{+}m^{+}(\theta) \int d\theta'
\sin{\Delta\theta} P_\textrm{c}(\Delta\theta) \; k(\theta')$, where $\Delta\theta = |\theta-\theta'|$ is the
collision angle and the geometrical factor $\sin{\Delta\theta}$ takes care of the collisional cross-section the
density of other microtubules present to the incoming one. Zippering events cause growing microtubule plus ends to
change direction, converting previously growing segments to the inactive state at a rate $\Phi
_{\text{\textit{zip}}}[m^{+},k]=v^{+}m^{+}(\theta) \int d\theta' \sin{\Delta\theta} P_\textrm{z}(\Delta\theta) \;
k(\theta')$. Simultaneously, new growing segments with an index $i+1$ are created, which is represented by the
boundary condition $m^{+}_{i+1}(l=0, \theta) = \int dl' \Phi _{\text{\textit{zip}}}[m^{+}_{i},k]$. This set of
boundary conditions is completed by a separate equation for $i=1$, which represents the isotropic nucleation of new
microtubules: $v^{+}m_{1}^{+}(l=0, \theta) =r_{\textrm{n}}/(2\pi)$. Finally, when a segment shrinks back to the point where
it had undergone a zippering event in the past, a previously inactive segment can be reactivated into a shrinking
state. Here we will not discuss the details of the rate $\Phi_{\text{\textit{react.}}}[m_{i}^{+},m_{i+1}^{-},k]$,
which contains a non-trivial history-dependence as a microtubule segment must ``un-zipper'' in the same direction
the zippering segment originally came from. We simply note that in the steady state Eq.\ (\ref{eq:MEinactive})
requires that this rate is balanced by the zippering rate discussed above.

In the steady state, the infinite set of equations \eqref{eq:MEall}
with the boundary conditions reduces to a set of four
coupled non-linear integral equations. These relate the length
density $k(\theta)$ to the average segment length, active segment
density and ratio between inactive and active segments, each being a
function of the angle $\theta$. It follows
that, for given interaction probabilities $P_c(\theta)$ and
$P_z(\theta)$, the remaining parameters can be absorbed into a
single dimensionless control parameter $G$, defined as
\begin{equation}
G = \left[ \frac{2 v^+ v^-}{r_{\textrm{n}}\left( v^+ + v^-
\right)}\right] ^{\frac{1}{3}}\left(
\frac{r_{\textrm{r}}}{v^{-}}-\frac{r_{\textrm{c}}}{v^{+}}
\right).\label{eq:Gdef}
\end{equation}
Here we only consider the case $G<0$, for which the
length of the microtubules is intrinsically bounded even in the
absence of collisions. In this case, the average length of
\emph{non-interacting} microtubules is given by $\bar{l}=(r_c/v_+ -
r_r/v^-)^{-1}$ \cite{Marileen93} and the control parameter
$G$ can be interpreted as $G=- l_0/\bar{l}$, implicitly defining an
interaction length scale $l_0$. As $G$ increases towards 0, the
number of interactions between microtubules increases.

For any value of $G$ there exists an isotropic solution to \eqref{eq:MEall}, for which the total length density
$\rho = \int \! \mathrm{d}\theta \, k(\theta)$  satisfies
$
l_0 \rho \left( \hat{c}_{0} l_0 \rho -2 G\right)^{2}=8,
$
where $\hat{c}_{n}$ denotes the $n$-th Fourier cosine coefficient
of the product $P_c(\theta) \left| \sin{\theta} \right|$. The
isotropic length density is therefore an increasing function of
the control parameter $G$ that only depends on the induced
catastrophes, and not on the probability of zippering. This can be
understood by the fact that zippering only serves to reorient the
microtubules, which has no net effect in the isotropic state.
Although a stationary isotropic solution exists for all values of
$G$, this solution is only stable for large negative values of
$G$. As $G$ increases, the number of interactions between
microtubules increases, until the isotropic solution becomes
unstable. This happens at the bifurcation point $G=G^*$, given by
\begin{align}
G^* &= (-2\hat{c}_2)^{1/3}\left( \frac{\hat{c}_0}{-2\hat{c}_2}
-1\right)\label{eq:bifG}.
\end{align}
We note that the location of the bifurcation point is determined solely by the properties of the induced catastrophe
probability $P_c(\theta)$, and, like the density in the isotropic phase, does not depend on zippering.

\begin{figure}[ht!]
\begin{center}
\includegraphics[width=0.4\textwidth]{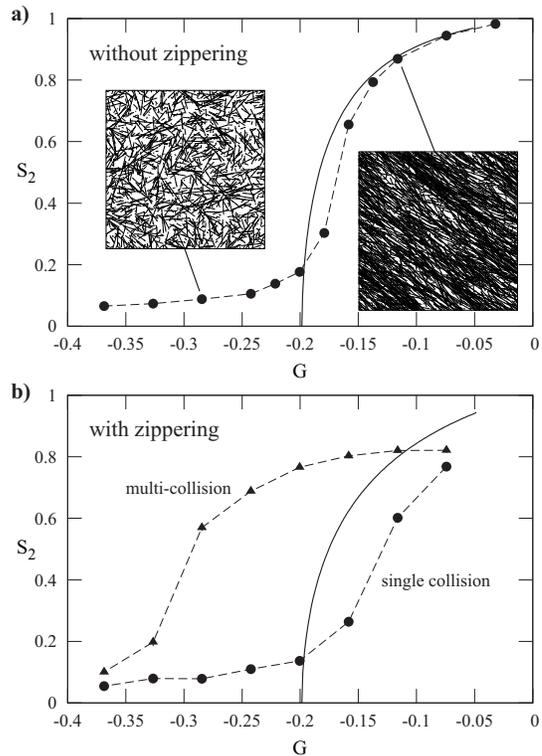}
\caption{Comparison between theoretical (solid lines)
and simulation results (symbols). The simulations were performed
on a 80 $\mu$m $\times$ 80 $\mu$m system with periodic boundary
conditions. The spontaneous catastrophe rate was varied to probe
different values of $G$: $r_c \in [4\times 10^{-3}$, $1.2\times
10^{-2}]$ s$^{-1}$. The nucleation rate was set to
$r_n=0.003\,\mu$m$^{-2}$s$^{-1}$ and other parameters were taken
from \cite{Vos+04} (interphase BY-2 cells): $v^{+} =$ 0.078 $\mu$m
s$^{-1}$, $v^{-}=$ 0.164 $\mu$m s$^{-1}$, $r_{r} =$ 6.8 $\times
10^{-3}$ s$^{-1}$. Measurements were performed after equilibrating
for 50,000s (a) or 250,000s (b); $G$ was increased between
measurements. The standard error of the mean is typically smaller
than the symbols and is otherwise indicated by vertical bars.
$N=$80(a),40(b).}\label{figure3}
\end{center}
\end{figure}

To quantify the degree of alignment we use the standard 2D nematic
order parameter $S_2$, defined as $S_2 = |\integral{\mathrm{e}^{i
2
\theta}k(\theta)}{\theta}{0}{2\pi}|/\integral{k(\theta)}{\theta}{0}{2\pi}$.
The full bifurcation diagram can be computed by numerically
tracing the ordered solution branch from the bifurcation point,
provided that the products $|\sin{\theta}| P_c(\theta)$ and
$|\sin{\theta}| P_z(\theta)$ have finite Fourier expansions. We
restrict ourselves to an expansion up to $\cos{4 \theta}$. The
coefficients are constrained by $\sin{(0)}P_c(0)=0$ and
$\sin{(0)}P_z(0)=0$. In line with experimental observations
\cite{DixCyr04} we choose the remaining parameters such that
$P_c(\theta)$ is monotonically increasing to a maximum at
$\theta=\pi/2$ and is maximally biased towards steep collision
angles (see \cite{Hawkins09} for other choices), and
$P_z(\pi/2)=0$. The magnitudes of $P_c(\theta)$ and $P_z(\theta)$
is similar to that observed in experiments, and the crossover
probability is fixed by the requirement
$P_c(\theta)+P_z(\theta)+P_x(\theta)=1$. The resulting interaction
probabilities are illustrated in Fig.\ \ref{figure2}b. We argue
that the apparent discrepancy with experiments, caused by setting
$P_z(0)=0$, is not very significant for the ordering transition,
as collisions between near-parallel microtubules are infrequent
and cause only slight changes of orientation in the case they lead
to zippering.

Given our choice for $P_c(\theta)$, we have $\hat{c}_0=3/8$ and
$\hat{c}_2=-1/4$ so that $G^* \approx -0.2$. The results are
representative for a large class of interaction probabilities with
$G^* < 0$. Higher modes do not affect the bifurcation point
\eqref{eq:bifG} and appear to have only minor effects on the
bifurcation diagram. Also, any changes to the overall magnitude of
$P_c(\theta)$ and $P_z(\theta)$ result only in a scaling of the
$G$-axis. Comparing the computed solutions (solid lines) for
systems with (Fig.\ \ref{figure3}b) and without (\ref{figure3}a)
zippering, we note that zippering has only a minor effect on the
ordering beyond the bifurcation point (see also \cite{Hawkins09}).
This shows that the `weeding out' of microtubules in the minority
direction through induced catastrophes is by itself sufficient to
explain microtubule alignment.

In parallel with the coarse-grained theoretical approach described above, we performed stochastic
particle-based simulations of the interacting microtubules. Fig.\ \ref{figure3} shows the resulting steady-state
alignment as a function of $G$, for systems with and without zippering. In the simulations, the presence of
zippering triggers the formation of microtubule bundles, in which aligned microtubules colocalize. In this case, we
need to specify how the interaction probabilities $P_c(\theta)$, $P_z(\theta)$ and $P_x(\theta)$ depend on the
number of microtubules that are present in both the incoming and encountered bundles. We investigate two extreme
scenarios. In the first scenario (single collisions) a microtubule treats a collision with a bundle as a single
collision, disregarding the other microtubules in both bundles. In the other scenario (multi-collisions) we
implicitly construct an effective interaction by sampling from the distribution of all multiple collisions and their
outcomes that can occur between an arbitrary microtubule from an incoming bundle with the full set of microtubules
in the target bundle (see Fig.\ \ref{figure2}a).

In the absence of zippering Fig.\ \ref{figure3}a, shows that the theoretical predictions and simulation results
agree well. As expected, the
agreement is less good when zippering is enabled (Fig.\ \ref{figure3}b), because zippering leads to strong spatial
correlations in the form of microtubule bundles, which are not accounted for in our mean-field-like theory. In the
case of the `multi-collision' interactions, the simulations indicate a significantly larger tendency to align,
whereas the system is less likely to align with `single' interactions. However, in both cases the behavior remains
qualitatively the same as the theoretical prediction and the alignment occurs over a similar range of $G$ values.

Finally we investigated the limit of weak interactions ($P_c(\theta), P_z(\theta) \ll 1$; data not shown) in which
the discrepancies due to the mean-field nature of our model should decrease. Without zippering
simulation results rapidly converge to the theoretical predictions. In the presence of zippering the results for the
`single' interactions deviate more strongly from the theory, because only a single collision is registered when a
microtubule encounters a bundle, effectively decreasing the density of interactions. The `multi-collision'
interaction however effectively accounts for the bundling, so that for progressively weaker interactions the
transition between the isotropic and ordered states converges to the predicted bifurcation point.

Our model of interacting cortical microtubules 
displays both isotropic and aligned phases
and is based on experimentally observed microscopic effects.
The kinetic parameters appearing in the control parameter $G$ may be regulated by the cell via MAPs, suggesting a mechanism for
cellular control over creation, maintenance and suppression of microtubule alignment.
Our results indicate that
collision-induced microtubule catastrophes alone could 
establish alignment in the cortical array of
plant cells. To what extent other known effects, such as microtubule treadmilling and severing, influence this mechanism, is a question we are currently addressing.

\begin{acknowledgments}
We thank Kostya Shundyak, Jan Vos and Jelmer Lindeboom for helpful
discussions. SHT was supported by the NWO programme
``Computational Life Sciences'' (Contract: CLS 635.100.003). RJH
was supported by the EU Network of Excellence ``Active Biomics''
(Contract: NMP4-CT-2004-516989). This work is part of the research
program of the ``Stichting voor Fundamenteel Onderzoek der Materie
(FOM)'', which is financially supported by the ``Nederlandse
organisatie voor Wetenschappelijk Onderzoek (NWO)''.
\end{acknowledgments}

\end{document}